\documentclass{llncs} 
\usepackage[dvipsnames]{xcolor} 
\usepackage{graphicx} 
\usepackage{dingbat}
\usepackage{svg} 
\usepackage{subcaption}
\usepackage{setspace}
\usepackage{comment}
\usepackage{blindtext}
\usepackage{enumitem}
\setlist{nolistsep}
\usepackage[toc,page]{appendix}
\usepackage{amsmath, amsfonts, amssymb}
\usepackage{graphicx}
\usepackage{rotating}
\usepackage{tabularx}
\newtheorem{insight}{Insight}
\newcommand{\ignore}[1]{}
\AtBeginEnvironment{tabular}{\singlespacing}

\begin{document}

\title{Characterizing the Evolution of Psychological Tactics  and Techniques Exploited by Malicious Emails}

\author{Theodore Longtchi 
 \and 
Shouhuai Xu
}

\institute{
Department of Computer Science\\ 
University of Colorado Colorado Springs\\
Colorado Springs, Colorado, USA
}

\maketitle

\begin{abstract}
The landscape of malicious emails and cyber social engineering attacks in general are constantly evolving. In order to design effective defenses against these attacks, we must deeply understand the Psychological Tactics (PTacs) and Psychological Techniques (PTechs) that are exploited by these attacks. In this paper we present a methodology for characterizing the evolution of PTacs and PTechs exploited by malicious emails. As a case study, we apply the methodology to a real-world dataset. This leads to a number insights, such as which PTacs (PTechs) are more often exploited than others. These insights shed light on directions for future research towards designing psychologically-principled solutions to effectively counter malicious emails. 
\end{abstract}

\keywords{Malicious Email \and Cyber Social Engineering Attack \and Psychological Tactic (PTac) \and Psychological Technique (PTech) \and Psychological Factor (PF)}

\section{Introduction}
Malicious emails are one major form of cyber social engineering attacks. For example, the 2023 Anti-Phishing Working Group (APWG) report \cite{apwg2023report} shows that the number of phishing attacks had tripled since May 2020 
and that more than one million domain names were used for phishing purposes in the year of 2023.
These alarming facts highlight that malicious emails and cyber social engineering attacks in general remain to be a rampart cyber threat. This is somewhat ironic because there have been many studies on designing defenses against these attacks, suggesting that existing defenses have achieved very limited success. Thus, it is imperative to understand the threat landscape.
In particular, this has motivated studies to understand the psychological aspects of cyber social engineering attacks, including malicious emails (e.g., \cite{montanez2023quantifying,khonji2013phishing,steves2019phish}).


In term of understanding psychological aspects of cyber social engineering attacks, three concepts have been investigated. One concept is known as
Psychological Tactic (PTac) \cite{PF-paper,montanez2023quantifying,montanez2022csekc}, which describes the overall deliberate thoughtfulness (and objective in a sense) of attackers in framing, for example, the malicious content of an email to influence an email recipient.
Another concept is 
Psychological Technique (PTech) \cite{PF-paper,montanez2023quantifying,montanez2022csekc}, which describes the psychologically relevant textual and imagery elements (e.g.) in email contents. The final of the three concepts is Psychological Factor (PF) \cite{PF-paper}, which describes an individual's psychological attributes or characteristics that may be exploited by cyber social engineering attacks. 
Studies, such as \cite{PF-paper,longtchi2024internet,vishwanath2011people,canham2022planting}, show that these psychological concepts have not been adequately considered in existing defenses, explaining their ineffectiveness. This motivates the present study to deepen our understanding of the exploitation PTacs and PTechs by malicious emails.

\noindent{\bf Our Contributions}. 
In this paper we make three contributions. 
First, we propose a methodology for characterizing the evolution of PTacs and PTechs exploited by malicious emails. The methodology includes the task of identifying the PTacs and PTechs exploited by malicious emails.
The methodology can be applied to any dataset of malicious emails to characterize the evolution of PTacs and PTechs exploited by these emails.
Second, we conduct a case study by applying the methodology to a real-world dataset of 1,260 malicious emails over two decades (2004-2024), with 60 malicious emails per year.
The case study leads to a number of insights, including: (i) the exploitation of PTacs and PTechs is potentially greatly affected by the emergence of major events such as the COVID pandemic; (ii) future defenses should pay more attention on countering the PTacs and PTechs that are often exploited by attackers, as we have yet to observe that attackers are forced to change their PTacs and/or PTechs; (iii) future defenses should strive to deal with the PTacs and PTechs that are often exploited together.
These insights shed light on promising research directions.

\noindent{\bf Ethical Issue}. After consulting with our institution's 
Internal Review Board (IRB), it is determined that no IRB approval is needed because the emails we analyze are from third parties (except for 108 malicious emails that are collected from our own email box).

\noindent{\bf Related Work}. 
There are many studies on cyber social engineering attacks, such as \cite{asiri2023survey,zieni2023phishing,syafitri2022social,alharbi2021social,PF-paper,chanti2020classification,aleroud2017phishing,wang2021social,frauenstein2020susceptibility,ferreira2015analysis,van2019cognitive,schaab2017social,ferreira2015principles,algarni2017empirical,kearney2016can,williams2017individual,goel2018mobile,cidon2019high} and the references therein.
The most closely related prior studies are: \cite{PF-paper}, which studied the evolution of PFs exploited by malicious emails; 
\cite{longtchi2024internet}, which presented the first systematization on 
the 16 PTechs and 46 PFs that have been exploited by cyber social engineering attacks;
and \cite{montanez2023quantifying}, which analyzed 7 PTacs and 8 PTechs exploited by 1,036 malicious emails.
However, these studies did not consider the evolution of PTacs or PTechs exploited by malicious emails. 
By contrast, we investigate the evolution of PTacs or PTechs in the span of 2004-2024.

Loosely related prior studies include: De Bona and Paci \cite{de2020real} showed, based on 
191 participants, that urgency is more likely to make employees susceptible to phishing attacks than authority; 
Gallo et al. \cite{gallo20212} studied how cognitive vulnerabilities are exploited by phishing emails based on a dataset of 2-year span; Wang et al. \cite{wang2021social} investigated susceptibility to phishing emails via visceral triggers; Gallo et al. \cite{gallo2024human} designed a system to detect persuasive elements in phishing email;
\cite{frauenstein2020susceptibility,ferreira2015analysis,van2019cognitive,ferreira2015principles} investigated how PTechs were exploited by malicious emails; and \cite{cidon2019high} leveraged PTechs to detect malicious emails. Each of these studies focused on one or very few PTechs. 
By contrast, we consider the evolution of 7 PTacs and 9 PTechs over two decades.


\smallskip

\noindent{\bf Paper Outline}.
Section \ref{sec:PTac-PTech_review} reviews and refines PTacs and PTechs.
Section \ref{sec:methodology} presents our methodology.
Section \ref{sec:case-study} describes a case study based on 1,260 malicious emails. Section \ref{sec:limitations} discusses our limitations. Section \ref{sec:conclusion} concludes the paper.

\section{Reviewing and Refining PTacs and PTechs}\label{sec:PTac-PTech_review}

The concepts of PTac and PTech were inspired by the concepts of Tactic and Technique in the MITRE ATT\&CK framework \cite{mitreattack}, but in the setting of cyber social engineering attacks \cite{PF-paper,montanez2023quantifying,montanez2022csekc}.
At a high level, a PTac can exploit one or multiple PTechs, and a PTech can exploit one or multiple PFs; 
PFs represent root causes of humans' susceptibility to cyber social engineering attacks including malicious emails. 
The investigation of the evolution of PFs is an independent work \cite{PF-paper}, where 20 PFs are reconciled from the 46 PFs proposed in \cite{longtchi2024internet}. 

\subsection{Reviewing the Concept of PTac}
\begin{definition}[PTac \cite{montanez2023quantifying,montanez2022csekc}]
In the context of malicious emails, a PTac describes the overall deliberate thoughtfulness (and objective) of the attacker in framing the malicious email content in order to victimize an email recipient.  
\end{definition}

PTacs can be seen as a measure of the attacker's effort to craft a malicious email with respect to an objective. 
There are 7 PTacs \cite{montanez2023quantifying}, all of which will be considered in this study.
(i) \texttt{Familiarity}, which refers to how an attacker attempts to establish a positive/trust relationship or association with a recipient (of a malicious email)  \cite{montanez2022social,alhamar2010ieee}.
(ii) \texttt{Immediacy}, which refers to the amplification of a time constraint to reduce a recipient's skepticism or scrutiny \cite{montanez2020human,nelms2016usenix}.
(iii) \texttt{Reward}, which refers to an exchange of something (physical or social) that is valuable to a recipient \cite{goel2017got,longtchi2022sok}.
(iv) \texttt{Threat of Loss}, which refers to an appeal to the recipient's desire to maintain a certain status and/or prevent losing an opportunity \cite{goel2017got,stajano2011understanding}.
(v) \texttt{Threat to Identity}, which refers to an attacker's efforts at manipulating a recipient's desire to maintain a positive and/or socially valuable reputation \cite{stajano2011understanding,montanez2022social}. 
(vi) \texttt{Claim to Legitimate Authority}, which refers to the exploitation of a legitimate power to obscure or deter a recipient's scrutiny \cite{stajano2011understanding,ferreira2015analysis}.
(vii) \texttt{Fit \& Form}, which refers to an attacker's effort at mimicking the expected style of authentic emails in a malicious email \cite{rajivan2018creative,goel2017got}.

\ignore{
Building from the findings in \cite{montanez2023quantifying}, the \textsl{Threat to Identity} PTac is the least exploited PTac in malicious emails. We could have refined the 7 PTacs to 6 PTacs by eliminating \textsl{Threat to Identity}, but we kept all 7 PTacs so that we could also determine this finding with respect to \textsl{Threat to Identity} using a new dataset of malicious emails. Therefore, we leverage all 7 PTacs in the current study to characterize the evolution of malicious emails. 
}

\subsection{Reviewing and Refining the Concept of PTech}

\begin{definition}[PTech \cite{PF-paper,montanez2023quantifying,montanez2022csekc}]
In the context of malicious emails, a PTech describes the psychologically relevant visual (textual and imagery) elements that an attacker employs in an email message to lure a recipient of the email. 
\end{definition}

PTechs can be seen as the visual elements exploited by a malicious email, such as highlighted texts or images that would have a psychological significance to a recipient to act as intended by the attacker. There are the 16 PTechs \cite{PF-paper}. 
For the purpose of the present study, we propose focusing on the PTechs that meet the following criteria: 
(i) A PTech must be distinct from the other PTechs as we observe that some of the 16 existing PTechs may overlap with each other.
(ii) A PTech should be exploited in a single email interaction to lure a victim rather than multiple interactions. This is a practical consideration when one cannot obtain emails of multiple interactions between an attacker and a recipient. 

The preceding criteria guide us to focus on the following 9 PTechs (out of the 16 PTechs presented in \cite{longtchi2024internet}), while noting that 8 (out of the 16) PTechs are considered in \cite{montanez2023quantifying}.
(i) \textsf{Persuasion}, which is the use of arguments pertaining to Cialdini's 6 Principles of Persuasion (i.e., authority, reciprocity, liking, scarcity, social proof, consistency / commitment) to convince a recipient \cite{frauenstein2020susceptibility,ferreira2015analysis}.
(ii) \textsf{Pretexting}, which is the use of a made-up story (or pretext) to justify contacting a recipient to gain the recipient's trust \cite{alhamar2010ieee,goel2017got}. 
(iii) \textsf{Impersonation}, which is the use of a false (and sometimes known) persona / entity to build trust with a recipient \cite{ferreira2015analysis,allodi2019need}.
(iv) \textsf{Visual Deception}, which is the use of visual but deceptive elements to gain the trust of a recipient
\cite{moreno2017fishing,montanez2020human}.
(v) \textsf{Incentive \& Motivator}, which is the use of textual elements to show financial benefit or gain to a  recipient  \cite{beckmann2018motivation,montanez2022csekc}.
(vi) \textsf{Urgency}, which is the use of textual elements to show a time constraint to urge a recipient to act quickly \cite{chowdhury2019impact,vishwanath2011people}. 
(vii) \textsf{Attention Grabbing}, which is the use of graphics or text to draw a recipient's attention \cite{nelms2016usenix,flores2015investigating}.
(viii) \textsf{Personalization}, which is about addressing a recipient by name or some personal identifiable information \cite{hirsh2012personalized,jagatic2007acm}.
(iv) \textsf{Contextualization}, which is the use of current or ongoing events (e.g., Covid-19) to 
engage a recipient \cite{XuIEEEISI2020-COVID19-Landscape,goel2017got}.


\section{Methodology}\label{sec:methodology}

The methodology has three steps: preparing a dataset;  identifying PTacs and PTechs exploited by malicious emails while noting the identification of PFs is described in \cite{PF-paper}; analysis.

\subsection{Preparing Dataset} 

We use the following guidance to help prepare datasets. (i) Researchers should determine the scope of their studies, including specifying the kinds of attacks they plan to study. This is relevant because there are different kinds of malicious emails, such as spear phishing vs. general phishing.
(ii) Researchers should determine the time span for collecting malicious emails. In principle, the longer the time span, the better. Moreover, researchers should determine the time granularity unit, such as year vs. month vs. week vs. day. These factors determine the effort made at analyzing a dataset, meaning that a feasible trade-off may need to be made. (iii) Researchers should assure the quality of malicious emails. This means that the sources of malicious emails should be trusted, and the malicious emails should be re-examined to confirm their maliciousness. 
Moreover, each malicious email should contain all of its original information / content. This is relevant because an email may contain some missing information, such as logos.
\vspace{-1.3em}
\subsection{Identifying PTacs and PTechs Exploited by Malicious Emails}\label{ss:id-PTacs-PTechs}

Inspired by (but different from) \cite{montanez2023quantifying}, we propose identifying or grading PTacs and PTechs from a given malicious email as follows.
We define the exploitation of a PTac or PTech based on the degree of application of the PTac or PTech in a malicious email: 0 for no application, 1 for implicit application, and 2 for explicit application. 
This allows us to analyze not only the absence (score 0) vs. presence (scores 1 and 2) of PFs, but also implicit vs. explicit exploitation (i.e., score 1 vs. score 2). The latter is important because implicit exploitation is stealthier than explicit exploitation and thus harder to defend against  because recognizing implicit exploitation of PTacs and PTechs would require understanding semantics of email contents, but recognizing explicit exploitation would only require recognition of their names (and variants) via keyword search.

We grade an email with respect to a PTac as follows: (i) We assign a score 0 to an email if the PTac is neither implicitly nor explicitly exploited by the email. For instance, an email containing ``\textit{Your account is scheduled to be suspended in 10 days"} would receive a score 0 with respect to the \texttt{Reward} PTac.
(ii) We assign a score 1 to an email if the PTac is implicitly exploited by the email. For instance, an email containing ``\textit{Your subscription has expired, update now"} would receive a score 1 with respect to the {\sf Immediacy} PTac because the word ``now" implies an implicit, but not explicit, exploitation of {\sf Immediacy}. (iii) We assign a score 2 to an email if the PTac is explicitly exploited by the email. For instance, an email containing ``\textit{Your reward is ready}'' would receive a score 2 with respect to the \texttt{Reward} PTac because the word ``reward'' is mentioned in the email.

Similarly, we grade an email with respect to a PTech as follows. 
(i) We assign a score 0 to an email if the PTech is neither implicitly nor explicitly exploited in the email. For instance, an email containing ``\textit{I've sent you the form. Please fill it out and send to me whenever you are ready.}'' would receive a score 0 with respect to the {\sf Urgency} PTech because the recipient has the control over the time of response.
(ii) We assign a score 1 to an email if elements of the PTech are implicitly, but not explicitly, exploited in the email; for instance, an email containing ``\textit{Download the file now}'' would receive a score 1 with respect to the {\sf Urgency} PTech  because the word ``now" signifies a time constraint on a recipient of the email.  
(iii) We assign a score 2 to an email if the PTech is explicitly exploited by the email. For instance, an email containing ``\textit{Treat this as a matter of urgency}'' would receive a score 2 with respect to the {\sf Urgency} PTech because of the word ``urgency'' is mentioned in the email.

\subsection{Analysis}
The analysis is centered at characterizing the evolution of PTacs and PTechs exploited by malicious emails. An analysis can be centered at answering many Research Questions (RQs), such as the following.

\begin{itemize}
\item RQ1: How frequently are PTacs and PTechs exploited by malicious emails?
Answering this question would allow us to draw insights into if PTacs (PTechs) are equally frequently exploited by malicious emails or not.
\item RQ2: Which PTacs and PTechs have been increasingly, decreasingly, or constantly exploited? Answering this question would allow us to draw insights into if attackers are forced to change their strategies in exploiting PTacs (PTechs) because the previously exploited PTacs (PTechs) are no longer effective (owing to effective defense).
\item RQ3: Which PTacs and/or PTechs are often exploited together? Answering this question would allow us to draw insights into thwarting the collective exploitation of PTacs and/or PTechs because it is neither wise nor feasible to design one defense against each PTac and/or PTech.

\item RQ4: What PTacs often exploit which PTechs, and what PTechs often exploit which PFs? Answering this question would allow us to draw insights into how defenses should be designed to thwart the collective exploitation of PTacs and PTechs (PTechs and PFs) because we cannot afford to design one defense against an individual pair of (PTac, PTech) or (PTech, PF).
\end{itemize}

\section{Case Study}\label{sec:case-study}

\subsection{Preparing Dataset}

The dataset is the same as the one analyzed in \cite{PF-paper}. That is, it is a dataset of malicious emails over the last 21 years (2004-2024). For each year, we select 60 emails, or 1,260 emails in total. The 1,260 emails are collected from various sources, including: the Anti-Phishing Working Group (APWG) (250), researchers (317), organizations (298), malicious emails reaching our own email box (108), and the Internet (287). The 1,260 emails are separated into 3 different types: phishing 74\%, scam 24\%, and spam 2\%. We manually checked the 1,260 emails and find that the spam emails are incorrectly labelled as phishing emails by their respective data sources.
We accomplish this by verifying the senders, while noting that many companies and institutions send out spam emails. 

\subsection{Identifying PTacs and PTechs from the 1,260 Malicious Emails}\label{ss:indentifying_PTac_PTech}

We identify or grade PTacs and PTechs from the 1,260 malicious emails as described in the methodology, while noting that the identification of PFs is described in \cite{PF-paper}. Owing to the amount of work and demand of expertise required for the study, we can only afford to have one PhD student grade the 1,260 emails.
To mitigate the potential inconsistency in between emails by the same grader (i.e., the PhD student), we use tables with examples of psychological elements or cues pertaining to the PTacs and PTechs. 

Table \ref{table:PFs_PTechs_PTacs_Cues}  is an excerpt of the real table that was used as assistant during the grading process.
Examples of psychological elements corresponding to each PF is presented in the second column, while noting that these elements provide evidence to the PTechs listed in column 3 and the PTacs listed in column 4. The reason for ordering the table as ``PF $\to$ Example elements or cues $\to$ PTech $\to$ PTac'' is that PTechs exploit PFs through elements or cues and PTacs exploit PTechs with the objective to lure recipients of an email. 

\vspace{-2em}
\begin{table}[!htbp]
{\small 
\begin{tabular}{|p{2cm} | p{4.8cm}| p{2.6cm}| p{2.2cm}|} 
\hline
\textbf{PF} & \textbf{Example elements or cues} & \textbf{PTech} & \textbf{PTac} \\
\hline
\hline
{\sc Impulsivity} &  ``Claim your price now" & {\sf Urgency}& {\tt Immediacy}\\ \hline
{\sc Trust} & University or Company logo & {\sf Visual Deception} & {\tt Familiarity} \\ \hline
{\sc Curiosity} & ``Congratulation! You won!" & {\sf Incentive \& Mot.} & {\tt Reward} \\ \hline
{\sc Cog. Miser} &``\textcolor{red}{Verify} your account" & {\sf Attention Grabbing} & {\tt Th. of Loss}\\ \hline
{\sc Ind. Diff.} &  ``Someone logged in your account" & {\sf Impersonation} & {\tt Familiarity} \\ \hline
{\sc Greed} & ``You have won \pounds 552,000.00..." & {\sf Incentive \& Mot.} & {\tt Reward} \\ \hline
{\sc Liking} &  ``We're committed to serving you" & {\sf Impersonation} & {\tt Le. Authority}\\ \hline
{\sc Laziness} & ``John, we're here for you." & {\sf Personalization} & {\tt Fit \& Form} \\ \hline
{\sc Sympathy} & ``She lost both parents in.." & {\sf Incentive \& Mot.} & {\tt Familiarity} \\ \hline
{\sc Vigilance} & ``G00GLE.COM" for Google.com & {\sf Visual Deception} &  {\tt Fit \& Form} \\ \hline 
{\sc Authority} & ``I'm Colonel...." / (Court logo ) & {\sf Impersonation} & {\tt Le. Authority} \\ \hline
{\sc Commitment} & Brand name (e.g., Walmart) & {\sf Persuasion} & {\tt Familiarity} \\ \hline
{\sc Soc. Proof} & ``As a member of our party" / & {\sf Contextualization} & {\tt Familiarity} \\ \hline
{\sc Expertise} &  (Technical of subject matter) & {\sf Contextualization} & {\tt Le. Authority} \\ \hline
{\sc {\scriptsize Defenselessness}} & ``Your account has been blocked"  & {\sf Urgency} & {\tt Immediacy} \\ \hline
{\sc Workload} &  ``Download and review it ASAP" & {\sf Urgency} & {\tt Immediacy} \\ \hline
{\sc Negligence} & (Hyperlink photo in an email) & {\sf Attention Grabbing} & {\tt Fit \& Form} \\ \hline
{\sc Scarcity} &  ``10 bottles for the first 10 callers" & {\sf Incentive \& Mot.} & {\tt Reward} \\ \hline
{\sc Loneliness} & ``Jasmine wants to meet you" & {\sf Pretexting} & {\tt Familiarity} \\ \hline
{\sc Reciprocity} &  ``...give back to the university..." & {\sf Persuasion} & {\tt Le. Authority} \\ \hline
\end{tabular}
}
\vspace{0em}
\caption{{\small Examples of PTechs exploiting PFs, where ``{\sc Cog. miser} stands for {\sc Cognitive miser}, ``Ind. diff.'' stands for {\sc Individual difference}, ``{\sc Soc. proof} stands for {\sc Social proof}, ``{\sf Incentive \& Mot.}'' stands for {\sf Incentive \& Motivator}, and ``{\tt Th. of Loss}'' stands for {\tt Threat of Loss}, ``{\tt Le. Authority}'' stands for {\tt Claim to Legitimate Authority}. 
}}
\label{table:PFs_PTechs_PTacs_Cues}
\end{table}
\vspace{-3em}

Figure \ref{fig:scam_email} presents an end-to-end example showing the PTacs, PTechs, and PFs that are exploited by a phishing email. In terms of grading PTacs, the email receives a score 1 with respect to the \texttt{Familiarity} PTac because it is implicitly exploited via the following elements: (i) the impersonation of a known company to make the email appear familiar to the recipient; and (ii) addressing the recipient by name to engender trust and familiarity with the attacker. The email receives a score 0 with respect to the \texttt{Fit \& Form} PTac because its composition failed in mimicking an email from McCarthy staffing company, namely the attacker's mistake in including the signature twice, which is a red flag alarm.

\begin{figure}[!htbp] 
\centering 
\includegraphics[width=\textwidth]{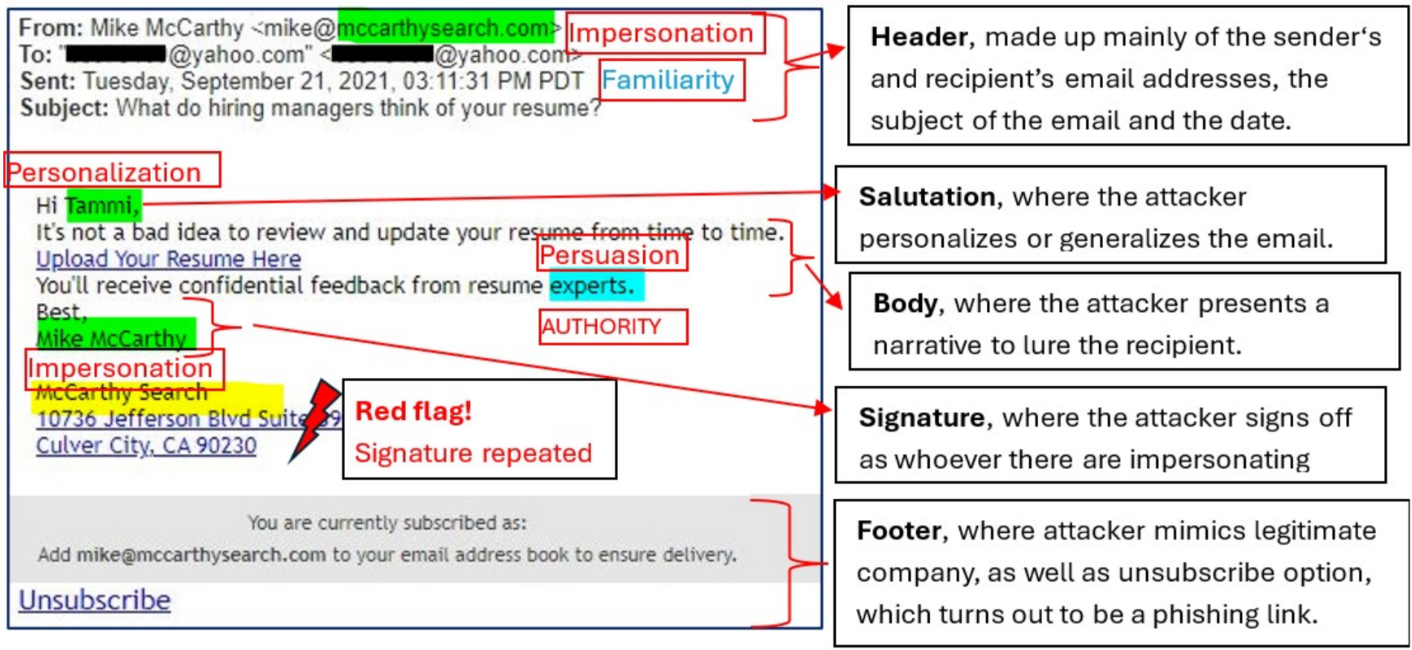}
\caption{\small A real-world phishing email showing which PTacs, PTechs, and PFs are exploited, where the exploited \texttt{Familiarity} PTac, {\sf Personalization} and {\sf Persuasion} PTechs, and {\sc authority} PF are highlighted. The email receives a score 0 with respect to the \texttt{Fit \& Form} PTac because the email has a mistake, which is the repeated Signature (i.e., an email from the McCarthy staffing company will not have this mistake, highlighted with the red flag).
}
\label{fig:scam_email}
\end{figure}

In terms of PTechs, the email receives a score 1 with respect to the {\sf Impersonation} PTech because it is implicitly exploited in the header of the email (highlighted in yellow), namely that the attacker impersonates the McCarthy Search Staffing company by spoofing its emails address. The email receives a score 1 with respect to {\sf Personalization} because it is implicitly exploited in the salutation, namely that the attacker addresses the recipient by name (i.e., ``Hi Tammi''). The email receives a score 1 with respect to the {\sf Persuasion} PTech because it is implicitly exploited via the ``expert" element that reflects the {\sc authority} PF \cite{PF-paper}, according to Cialdini's Principles of Persuasion \cite{ferreira2015principles}. 

\ignore{
\vspace{-2em}
\begin{figure}[!htbp] 
\centering 
\def\svgwidth{\columnwidth} 
\includesvg[inkscapelatex=false, width =\textwidth]{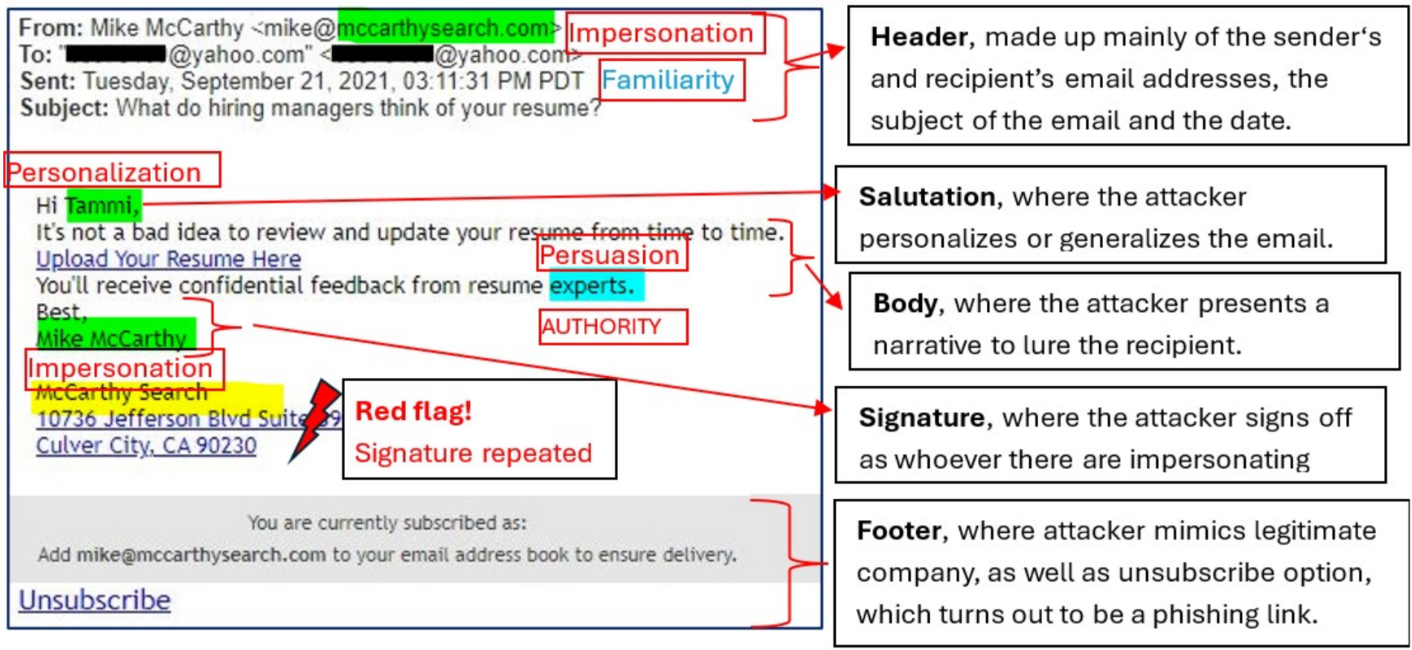}
\caption{\small 
A real-world phishing email showing which PTacs, PTechs, and PFs are exploited, where the exploited \texttt{Familiarity} PTac, {\sf Personalization} and {\sf Persuasion} PTechs, and {\sc authority} PF are highlighted. 
The email receives a score 0 with respect to the \texttt{Fit \& Form} PTac because the email has a mistake, which is the repeated Signature (i.e., an email from the McCarthy staffing company will not have this mistake, highlighted with the red flag).
}
\label{fig:scam_email}
\end{figure}
\vspace{-2em}
}

In terms of PFs, the email receives a score 1 with respect to the {\sc authority} PF because the word ``experts'' is used in the email to describe the email sender's expertise. Note that the email receives a score 0 with respect to the {\sc expertise} PF because the word ``experts" describes the expertise of the sender, rather than the expertise of the recipient.

\subsection{Analysis}

\ignore{

\begin{figure}[!htbp] 
\centering 
\def\svgwidth{\columnwidth} 
\includesvg[inkscapelatex=false, width =\textwidth]{images/Mapping_with_PTacs2.svg}
\caption{\small Mappings of relationships between PTacs, PTechs, and PFs, where a dashed line inside a box indicates different categories, a filled circle indicates PFs with empirical quantitative studies and an empty circle otherwise, the ``+'' (``-'') sign indicates that a factors increases (decreases) human susceptibility to attacks. ``Ind.'' is short for  Individual, and ``com.'' is short for commitment.}
\label{fig:mapping_PTac}
\end{figure}

}

\begin{sidewaystable}
\begingroup
\setlength{\tabcolsep}{3pt} 
\renewcommand{\arraystretch}{1.10} 
\scriptsize
\begin{tabular}{|l|c|c|c|c|c|c|c|c|c|c|c|c|c|c|c|c|c|c|c|c|c|c|}
 \hline
\textbf{PTac} & 2004&	2005&	2006&	2007&	2008&	2009&	2010&	2011&	2012&	2013&	2014&	2015&	2016&	2017&	2018&	2019&	2020&	2021&	2022&	2023&	2024&	TPTac \\ \hline
{\tt Familiarity}&	35&	21&	31&	19&	30&	23&	18&	20&	24&	31&	52&	46&	49&	47&	32&	35&	40&	33&	17&	41&	25&	669 \\ \hline
{\tt Immediacy}&	25&	15&	27&	20&	22&	12&	9&	16&	15&	20&	30&	26&	31&	23&	28&	32&	37&	22&	22&	32&	23&	487 \\ \hline
{\tt Reward}&	18&	39&	15&	30&	31&	30&	28&	19&	16&	17&	4&	7&	5&	7&	12&	10&	11&	21&	13&	4&	16&	353 \\ \hline
{\tt Threat of Loss}&	19&	4&	19&	13&	14&	13&	7&	13&	24&	23&	51&	39&	47&	35&	21&	25&	16&	20&	14&	28&	18&	463 \\ \hline
{\tt Threat to Identity}&	2&	1&	0&	1&	0&	0&	0&	0&	3&	0&	0&	2&	0&	0&	1&	2&	1&	1&	1&	1&	&	16 \\ \hline
{\tt Claim to Legit Authority}&	27&	23&	28&	27&	35&	26&	28&	25&	34&	36&	55&	47&	51&	52&	36&	39&	50&	42&	31&	42&	38&	772 \\ \hline
Fit \& Form&	33&	44&	42&	45&	50&	43&	50&	51&	46&	52&	52&	52&	54&	52&	44&	51&	57&	53&	42&	47&	44&	1004 \\ \hline
Total &	159&	147&	162&	155&	182&	147&	140&	144&	162&	179&	244&	219&	237&	216&	174&	194&	212&	192&	140&	195&	164& \textbf{3764} \\	\hline
	&&&&&&&&&&&&&&&&&&&&&& \\																						
\textbf{PTech} &	2004&	2005&	2006&	2007&	2008&	2009&	2010&	2011&	2012&	2013&	2014&	2015&	2016&	2017&	2018&	2019&	2020&	2021&	2022&	2023&	2024& TPTech \\ \hline
{\sf Urgency}&	24&	31&	32&	22&	15&	19&	17&	21&	13&	19&	22&	29&	32&	24&	28&	32&	32&	25&	18&	26&	24&	505 \\ \hline
{\sf Visual Deception}&	14&	5&	19&	11&	22&	3&	7&	13&	10&	21&	15&	6&	3&	8&	16&	18&	16&	20&	19&	25&	18&	289 \\ \hline
{\sf Incentive \& Motivator}&	17&	43&	14&	32&	31&	33&	28&	22&	18&	21&	5&	7&	5&	7&	16&	11&	13&	22&	21&	4&	20&	390 \\ \hline
{\sf Persuasion}&	11&	22&	17&	30&	25&	16&	8&	13&	17&	11&	18&	13&	18&	26&	9&	6&	22&	8&	9&	10&	6&	315 \\ \hline
{\sf Impersonation}&	41&	42&	41&	39&	50&	43&	41&	28&	35&	42&	56&	47&	56&	52&	39&	45&	56&	50&	38&	47&	43&	931 \\ \hline
{\sf Contextualization}&	8&	33&	16&	17&	25&	20&	26&	18&	10&	13&	1&	5&	3&	6&	11&	6&	32&	24&	12&	5&	11&	302 \\  \hline
{\sf Pretexting}&	32&	40&	36&	39&	37&	27&	15&	27&	24&	37&	49&	52&	46&	49&	30&	36&	36&	41&	32&	46&	41&	772 \\ \hline
{\sf Personalization}&	5&	5&	1&	4&	3&	0&	4&	12&	12&	9&	2&	0&	3&	7&	6&	5&	4&	15&	5&	10&	5&	117 \\ \hline
{\sf Attention Grabbing}&	26&	37&	53&	52&	55&	42&	57&	55&	55&	57&	50&	51&	57&	48&	46&	52&	31&	53&	48&	53&	52&	1030 \\ \hline
Total&	178&	258&	229&	246&	263&	203&	203&	209&	194&	230&	218&	210&	223&	227&	201&	211&	242&	258&	202&	226&	220& \textbf{4651} \\ \hline
&&&&&&&&&&&&&&&&&&&&&& \\																					
\textbf{PF} &	2004&	2005&	2006&	2007&	2008&	2009&	2010&	2011&	2012&	2013&	2014&	2015&	2016&	2017&	2018&	2019&	2020&	2021&	2022&	2023&	2024&	TPF \\ \hline
{\sc Individual Difference}&	15&	13&	11&	15&	9&	10&	13&	15&	16&	16&	10&	17&	16&	28&	18&	23&	16&	22&	24&	14&	15&	336 \\ \hline
{\sc Trust}&	25&	26&	16&	14&	19&	17&	22&	26&	31&	26&	50&	50&	40&	44&	26&	35&	38&	41&	27&	41&	32&	646 \\ \hline
{\sc Impulsivity}&	32&	36&	32&	45&	47&	40&	41&	28&	46&	50&	55&	54&	50&	52&	43&	49&	48&	54&	47&	57&	53&	959 \\ \hline
{\sc Vigilance}&	9&	11&	7&	0&	0&	0&	0&	2&	2&	0&	1&	0&	1&	0&	1&	0&	0&	0&	0&	0&	0&	34 \\ \hline
{\sc Greed}&	17&	26&	5&	17&	21&	26&	12&	10&	3&	9&	1&	1&	2&	1&	8&	2&	5&	8&	5&	1&	7&	187 \\ \hline 
{\sc Sympathy}&	1&	1&	0&	10&	10&	5&	3&	4&	2&	2&	0&	1&	0&	0&	0&	1&	0&	0&	5&	0&	0&	45 \\ \hline
{\sc Liking}&	3&	2&	0&	11&	7&	5&	3&	5&	3&	5&	3&	2&	4&	4&	11&	8&	13&	8&	3&	7&	6&	113 \\ \hline
{\sc Curiosity}& 	17&	33&	20&	39&	36&	40&	35&	30&	32&	31&	6&	13&	17&	17&	33&	28&	34&	33&	36&	31&	36&	597 \\ \hline
{\sc Laziness}&	3&	10&	4&	1&	1&	2&	5&	1&	2&	2&	0&	6&	0&	4&	5&	4&	0&	1&	0&	3&	0&	54 \\ \hline
{\sc Cognitive Miser}&	15&	14&	12&	6&	12&	8&	11&	10&	21&	18&	42&	37&	43&	42&	25&	31&	35&	23&	19&	24&	21&	469 \\ \hline
{\sc Social Proof}&	4&	2&	2&	2&	2&	3&	5&	5&	1&	0&	2&	1&	3&	1&	2&	3&	0&	4&	1&	0&	0&	43 \\ \hline
{\sc Authority}&	17&	5&	13&	17&	25&	13&	5&	14&	20&	21&	49&	46&	50&	43&	24&	28&	37&	20&	14&	25&	22&	508 \\ \hline
{\sc Expertise}&	0&	0&	1&	1&	1&	2&	1&	0&	0&	0&	2&	1&	0&	0&	0&	0&	0&	2&	0&	0&	0&	11 \\ \hline
{\sc Scarcity}&	0&	6&	8&	5&	2&	4&	3&	4&	3&	5&	17&	16&	1&	0&	2&	1&	0&	0&	0&	4&	0&	81 \\ \hline
{\sc Commitment}&	6&	0&	10&	13&	12&	11&	7&	11&	3&	8&	16&	9&	7&	14&	22&	20&	18&	16&	13&	18&	17&	251 \\ \hline
{\sc Negligence}&	7&	1&	0&	0&	1&	4&	0&	0&	5&	6&	3&	6&	7&	10&	11&	6&	4&	6&	7&	11&	6&	101 \\ \hline
{\sc Defenselessness}&	2&	3&	5&	2&	3&	1&	9&	6&	14&	11&	7&	7&	10&	9&	30&	20&	32&	26&	28&	30&	23&	278 \\ \hline
{\sc Loneliness}&	4&	4&	1&	4&	3&	8&	10&	8&	2&	6&	0&	1&	1&	0&	3&	0&	1&	2&	1&	0&	0&	59 \\ \hline
{\sc Workload}&	6&	1&	1&	2&	2&	2&	2&	0&	9&	5&	16&	15&	27&	29&	8&	13&	10&	13&	18&	16&	17&	212 \\ \hline
{\sc Reciprocity}&	1&	1&	0&	0&	0&	0&	0&	1&	0&	0&	1&	0&	0&	0&	1&	0&	0&	0&	0&	0&	0&	5 \\ \hline
Total&	184&	195&	148&	204&	213&	201&	187&	180&	215&	221&	281&	283&	279&	298&	273&	272&	291&	279&	248&	282&	255& \textbf{4989} \\	\hline 

\end{tabular}
\caption{{\scriptsize Counts of PTacs, PTechs, and PFs exploited by the 1,260 emails per year, as well as the  
Total of PTac (TPTac), the Total of PTech (TPTech), and the Total of PF (TPF) from 2004 to 2024.  
}}
\label{tab:Constructs_counts}
\endgroup
\end{sidewaystable}

Our analysis first considers the absence vs. presence of a PTac or PTech in an email, namely an email receiving a score of 0 vs. a score of 1 or 2 with respect to a PTac or PTech. We may further consider the distinction of score 1 vs. 2 (i.e., implicit vs. explicit exploitation) when the need arises, especially for the purpose of determining how evasive the attacks are.

\bigskip 

\noindent{\bf Addressing RQ1: How frequently are PTacs and PTechs exploited by malicious emails?} 
To answer this RQ, we only need to differentiate the absence vs. presence of a PTac or PTech, namely a score of 0 according to the grading method described in the methodology vs. a non-zero score (i.e., score 1 or 2 with respect to a PTac or PTech according to the grading method described in the methodology).
Table \ref{tab:Constructs_counts} presents the detailed results.

\begin{figure}[!htbp]
\begin{subfigure}[t]{0.5\textwidth}
\includegraphics[height=0.7\textwidth]{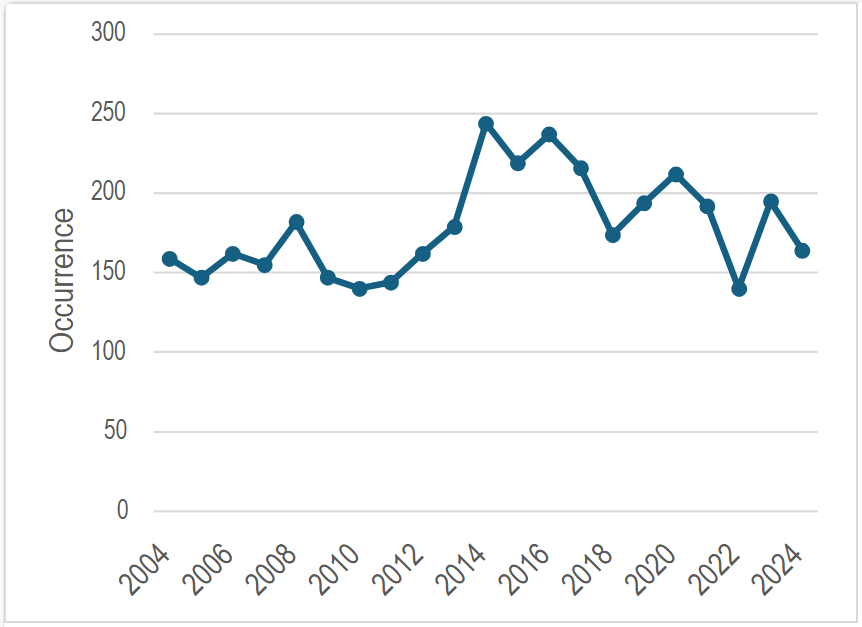}
  \caption{Total occurrence of PTacs}
  \label{fig:Occurrence_PTacs_All_BY}
\end{subfigure}
\hfill
\begin{subfigure}[t]{0.50\textwidth}
\includegraphics[height=0.7\textwidth]{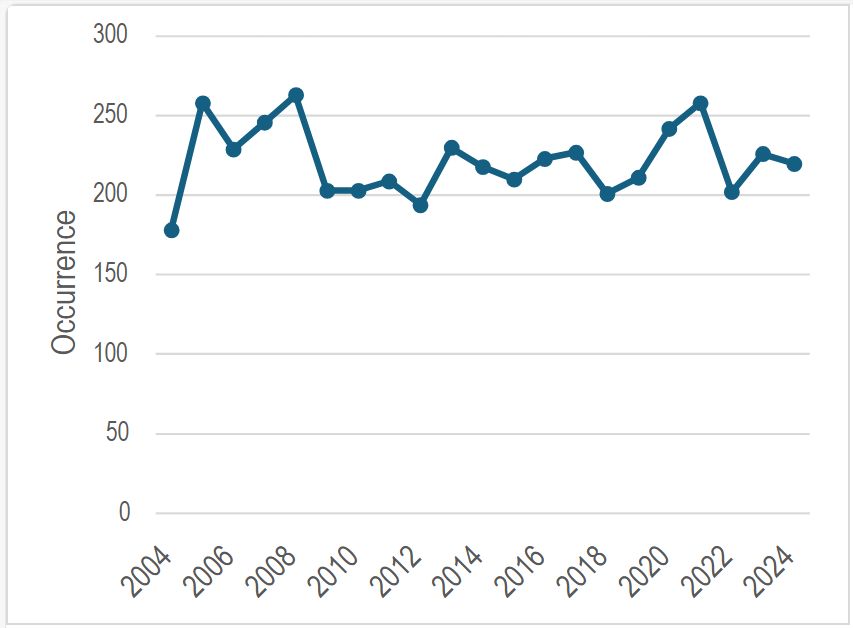}
  \caption{Total occurrence of PTechs}
  \label{fig:Occurrence_PTechs_All_BY}
\end{subfigure}
\vspace{-1em}
\caption{{\small Evolution of exploited PTacs and PTechs in malicious emails from 2004 to 2024}}
\label{fig:PTacs_PTechs_occurrences_aggregated}
\end{figure}

Figure 
\ref{fig:PTacs_PTechs_occurrences_aggregated}(a) plots the occurrence of all the PTacs over the 21 years, namely the aggregated frequency (i.e., the sum of the instances) of PTacs exploited by malicious emails, where occurrence in each year is upped bounded by 420 (recalling that we have 60 emails per year and 7 PTacs). 
We make the following 
observations. 
(i) The exploitation of PTacs has, by and large, somewhat increased over the 21 years. There is a surge in 2014, which is mainly caused by the surge in the exploitation of \texttt{Familiarity}, {\tt Threat of Loss}, and \texttt{Claim to Legitimate Authority}. By looking into the emails, we find that this is caused by surges in the impersonation of authorities, such as a bank or the U.S. Internal Revenue Service (IRS), in the impersonation of familiar entities, such as PayPal or university IT desk demanding an account update, and in the use of attacks containing messages like ``\textit{update your password or your account will be suspended}.'' 
(ii) Almost all the PTacs have been exploited by malicious emails on a yearly basis.
(iii) The \texttt{threat to identity} PTac has been rarely exploited over the 21 year.

Figure 
\ref{fig:PTacs_PTechs_occurrences_aggregated}(b)
plots the occurrence of all the PTechs over the 21 years, where the occurrence is upper bounded by 540 (as we have 60 emails per year and 9 PTechs).
We make the following observations. (i)
There is, by and large, a general trend that PTechs are increasingly exploited.
There is an explosive increase in 2005 
perhaps because the Nigerian Prince (or 419) scams became popular as 26 out of the 60 emails are of this kind. The drop in the exploitation of PTechs in 2009 to 2012 is caused by the drop in the exploitation of the {\sf Incentive \& Motivator}, {\sf Persuasion}, and {\sf Pretexting} PTechs. The increase in the exploitation of PTechs in 2021 may be attributed to the emergence of the Covid-19 pandemic whereby malicious emails attempted to scam people with Covid-19 related messages. This may be justified by the fact that 22 out of the 60 emails are in this category, while 8 emails explicitly use Covid-19 as a pretext and 14 emails use work from home offers during the Covid-19 lockdown. However, these increases and decrease could well be caused by the small sample size (i.e., 60 emails per year).
(ii) Almost all PTechs have been exploited on a yearly basis. 

\begin{insight}
\label{insight:PTacs-and-PTechs}
The exploitation of PTacs and PTechs over time is potentially largely affected by the emergence of major events that provide attacks with opportunities to wage new kinds of malicious emails. 
\end{insight}

Insight \ref{insight:PTacs-and-PTechs} resonates findings by other researchers in closely related contexts such as malicious websites (e.g., \cite{XuIEEEISI2020-Malicious-Websites,XuIEEEISI2020-COVID19-Landscape,MirIEEECNS2022}). Moreover, it stresses the importance in designing proactive defenses against cyber social engineering attacks, including but not limited to, malicious emails that exploit emerging events to wage new kinds of attacks. 

\smallskip

\begin{figure}[!htbp]
\centering
\includegraphics[height=0.42\textwidth]{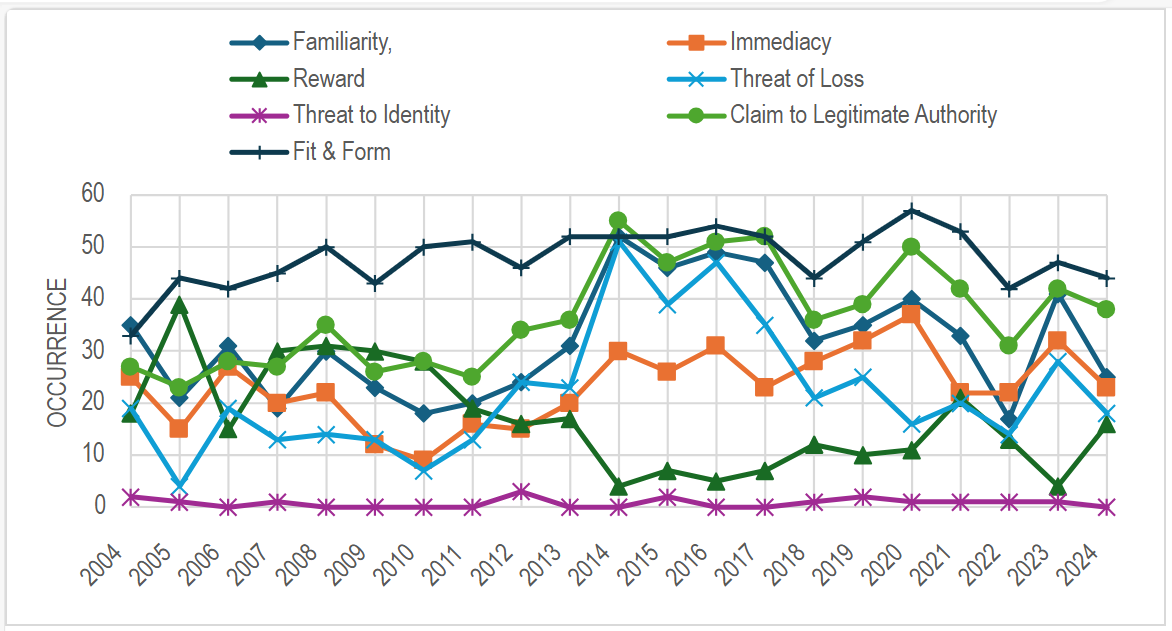}
\vspace{-.5em}
\caption{Evolution of the exploitation of individual PTacs}
\label{fig:Occurrence_PTacs_BY2}
\end{figure}

\noindent{\bf Addressing RQ2: Which PTacs and PTechs have been increasingly, decreasingly, or constantly exploited?}
Figure \ref{fig:Occurrence_PTacs_BY2}
plots the evolution of PTacs exploited by the 1,260 malicious emails over the last 21 years; each year, the number of occurrences (the $y$-axis) is the number of instances that a PTac is exploited by the 60 emails  (i.e., bounded from above by 60). 

\vspace{-2em}
\vspace{-2em}

We make the following observations. (i) \texttt{Fit \& Form} is the most exploited PTac and \texttt{Familiarity} is a frequently exploited PTac. This is different from the finding presented in \cite{montanez2023quantifying} that \texttt{Familiarity} was the most exploited PTac. The discrepancy can be attributed to the fact that the two studies used two different datasets. 
(ii) \texttt{Threat to identity} is the least exploited PTac over the 21 years, with a total of 16 instances in the 1,260 emails (i.e., 0 exploitations in some years). The second least exploited PTac is \texttt{Reward}, with a total of 353 instances in the 1,260 emails. 
(iii) \texttt{Reward} is the only PTac that shows a relatively decreasing exploitation from its peak in 2005 (39 instances in 60 emails) to its lowest exploitation in 2014 (4 instances in 60 emails) and remaining relatively low before rising again in 2021, then dropping again in 2022 and 2023. Although the sample (i.e., 60 emails) is too small to draw reliable conclusions, we speculate that this trend may be due to the gradual decline of scams from the early days of the Internet boom and the increase of Covid-19 relative scams that peaked in 2021 with stimulus check scams. (iv) The {\tt Immediacy} and \textit{Threat to Identity} PTacs have a relatively constant exploitation from 2004 to 2024. This suggests that the attackers see no need to change tactics that have been working  for them.

\begin{figure}[!htbp]
\centering
\includegraphics[height=0.42\textwidth]{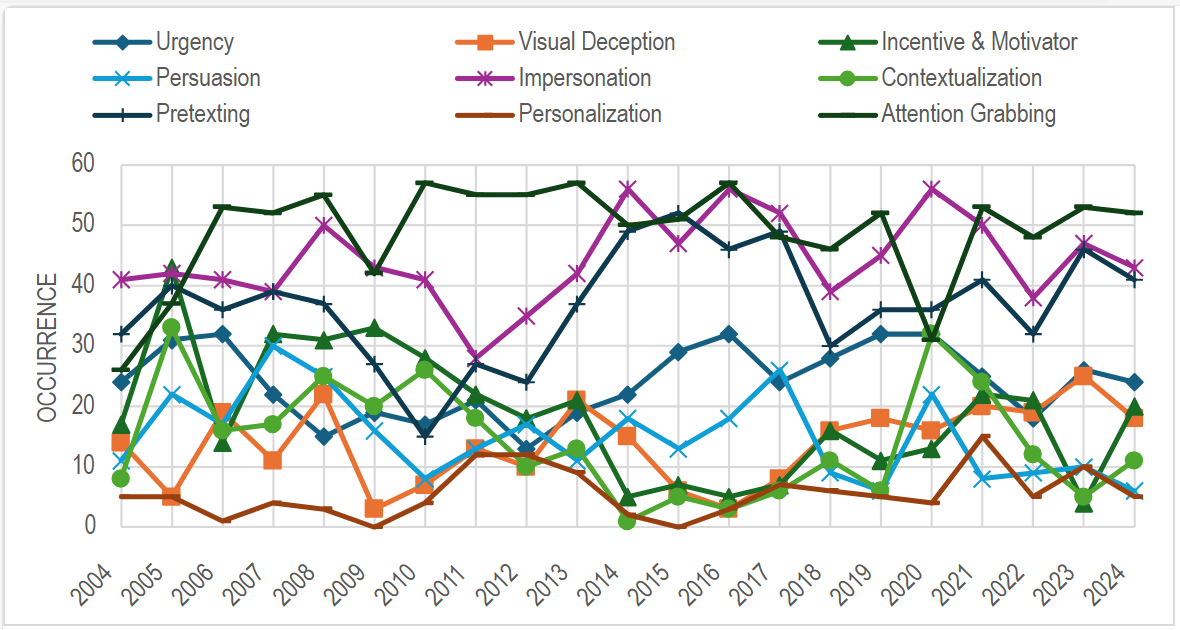}
\vspace{-.5em}
\caption{{\small Evolution of the exploitation of individual PTechs}}
\label{fig:PTechs_occurrences}
\end{figure}

Similarly, Figure \ref{fig:PTechs_occurrences} plots the evolution of PTechs exploited by the 1,260 malicious emails over the 21 years, where the number of occurrence each year is also bounded from above by 60 (i.e., 60 emails per year). 
We make the following observations. (i) {\sf Attention Grabbing} PTech is the most exploited PTech with a total of 1,030 instances in 1,260 emails, which is followed by {\sf Impersonation} with 931 instances and {\sf Pretexting} with 772 instances. (ii) {\sf Personalization} is the least exploited PTech with 117 instances in 1,260 emails. This may be attributed to some of the following reasons: it is difficult for attackers to get personal details of recipients in order to send personalized emails, attackers can be too lazy to carry out meaningful reconnaissance, some attackers may not be equipped with tools and skills to carry out reconnaissance, and the personalization of emails does not give attackers the leverage to send bulk of emails.
(iii) {\sf Visual Deception}, {\sf Pretexting}, {\sf Impersonation}, and {\sf Attention Grabbing} have all increased from 2004 to 2024. This may be attributed to attackers applying more visuals in malicious emails to grab recipients' attention and using false narratives, such as pretexts, to disarm recipients' suspicion.  
(iv) {\sf Incentive \& Motivator} has seen a relative decrease in exploitation from its peak in 2005 (43 instances in 60 emails) to 2014 (5 instances in 60 emails), then levels up to in 2017 before a bumpy rise in 2024 with 20 instances in 60 emails. This decrease may be attributed to the decrease of scam emails offering incentives, such as the Nigerian Prince scams, since they became known to the public and thus decreasingly successful.
(v) {\sf Urgency} and {\sf Personalization} have seen a relatively constant exploitation by malicious emails from 2004 to 2024. This may be attributed to the fact that attackers 
often employ time-constrained elements of {\sf Urgency} to force recipients to act without thoughtfulness.

\vspace{-2em}
\vspace{-2em}

\begin{insight}
Future defenses should pay more attention on coping with the PTacs (e.g., {\tt Fit \& Form}) and PTechs (e.g., {\sf Attention Grabbing}) that are more exploited than others, as we have yet to witness that attackers are forced to change their  PTacs and/or PTechs.
\end{insight}

\begin{figure}[!htbp] 
\centering 
\begin{subfigure}[t]{0.9\textwidth}
    \centering
    \includegraphics[width = .9\textwidth]{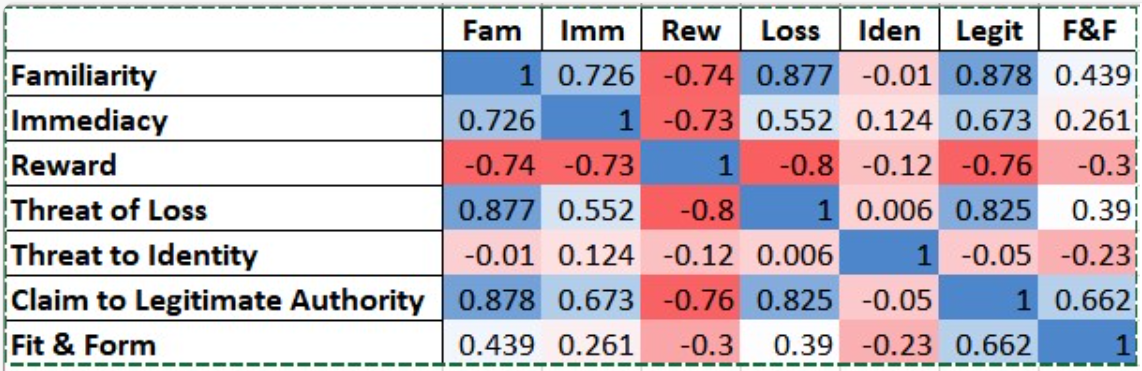}
    \caption{Correlation between PTacs and PTacs.}
    \label{fig:PTac-PTac_Corr}
\end{subfigure} 
\begin{subfigure}[t]{0.9\textwidth}
    \centering
    \includegraphics[width = .9\textwidth]{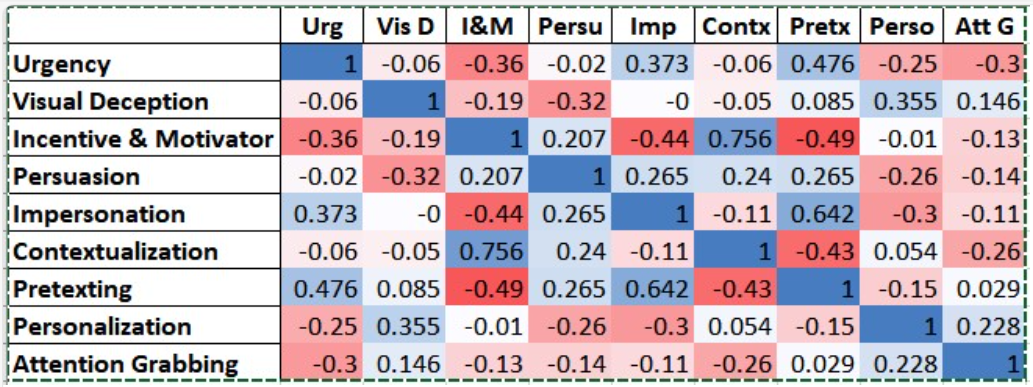}
    \caption{Correlation between PTechs and PTechcs.}
\label{fig:PTech-PTech_Corr}
\end{subfigure}
\begin{subfigure}[t]{0.9\textwidth}
    \centering
    \includegraphics[width = .9\textwidth]{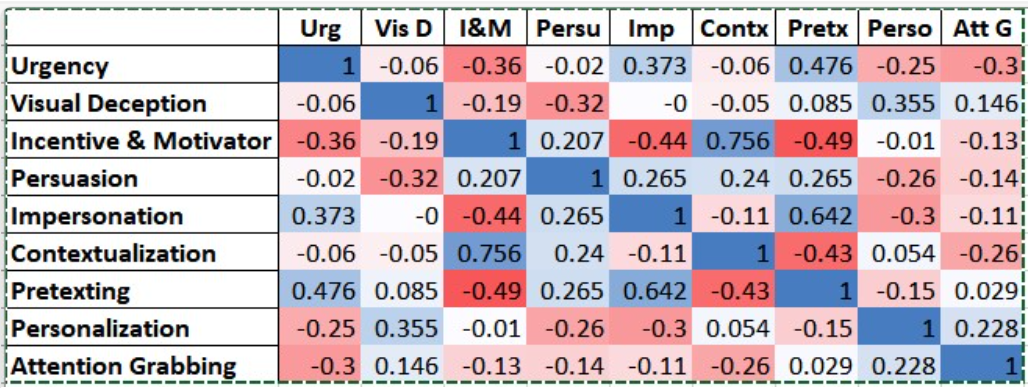}    
    \caption{Correlation between PTacs and PTechs.}
\label{fig:PTac-PTech_Corr}
\end{subfigure}
\vspace{-1em}
\caption{\small Correlations in color coding: positive correction increases from white to a depth of blue, and negative correction (absolute value) increase from white to a depth of red.} 
\label{fig:PTac_PTech_corr}
\end{figure}

\noindent{\bf Addressing RQ3: Which PTacs and/or PTechs are often exploited together?} To address this question, we use the Pearson correlation coefficient \cite{schober2018correlation}, denoted by $r$, to characterize the correlations between PTacs (i.e., PTac-PTac correlation), PTechs (i.e., PTech-PTech correlation), and PTacs and PTechs  (i.e., PTac-PTech correlation). These coefficients are computed by treating the 1,260 emails as a sample.
Figure \ref{fig:PTac_PTech_corr} presents the correlation coefficients, where intuitive abbreviations are used (e.g., ``Rew'' for {\tt Reward}). Elaborations follow.

\vspace{-2em}
\vspace{-2em}

For PTac-PTac correlations, we make the following observations from Figure \ref{fig:PTac_PTech_corr}(a).
(i) 
The strongest positive correlation is between \texttt{Familiarity} and \texttt{Claim to Legitimate Authority} with coefficient $r = 0.878$, perhaps because attackers usually impersonate entities of authority that are known to a recipient.
Another strong positive correlation occurs between \texttt{Familiarity} and \texttt{Immediacy} with $r = 0.826$, perhaps because attackers often impersonate an entity familiar to a recipient while urging the recipient to take quick actions. 
We use one example email from the dataset to illustrate these positive correlation: An attacker impersonates a university IT desk and sends an email stating, ``You've used 97\% of your email storage. You have to verify your email within 24 hours, or you will lose your email and your free storage." In this attack, the claimed university IT is familiar to the recipient (\texttt{Familiarity}); the attacker claims to be the university IT desk (\texttt{Claim to Legitimate Authority}); the recipient (e.g., a student) will lose their email storage if they do not act within 24 hours (\texttt{Immediacy}); and the claimed sender has the power to take the recipient's email away (\texttt{Threat of Loss}). 
(ii) The strongest negative correlation is between \texttt{Reward} and \texttt{Threat of Loss} with $r = -0.80$. This is natural because
attackers do not offer reward and incur loss to recipients in the same email.
Another strong negative correlation is between \texttt{Claim to Legitimacy} and \texttt{Reward} with $r = -0.76$. This also natural because a legitimate power would not offer reward to a recipient.

For PTech-PTech correlations, we make the following observations from Figure \ref{fig:PTac_PTech_corr}(b).
(i) 
The strongest positive correlation is between {\sf Contextualization} and {\sf Incentive \& Motivator} with $r = 0.756$ because attackers often leverage current events to create emails that offer incentives to the email recipients. For example, one email from the dataset states, ``\$2400 has been allocated to you as part of the Covid-19 aid stimulus bill", where Covid-19 pandemic is the context ({\sf Contextualization}) and the stimulus check is the incentive to motivate the recipient to act quickly ({\sf Incentive \& Motivator}). 
Another positive correlation is between {\sf Pretexting} and {\sf Impersonation} with $r = 0.642$, perhaps because attackers often need to impersonate a trusted entity in order to present the pretext for contacting a recipient. 
(ii) The strongest negative correlation is between {\sf Pretexting} and {\sf Incentive \& Motivator} with $r=-0.49$ because these PTechs should not be exploited in a single email.

For PTac-PTech correlations, we make the following observations from Figure \ref{fig:PTac_PTech_corr}(c).
(i) 
The strongest positive correlation  is between the \texttt{Reward} PTac and the {\sf Incentive \& Motivator} PTech with $r = 0.979$, perhaps because attackers usually use monetary reward as incentives to lure victims. Another strong positive correlation is between the \texttt{Claim to Legitimate Authority} PTac and the {\sf Impersonation} PTech with $r=0.815$, because attackers have to impersonate some authority.
Yet another strong positive correlation is between the \texttt{Familiarity} PTac and the {\sf Impersonation} PTech with $r=0.783$, because attackers always try to impersonate a personality or entity that is familiar to the recipient.
(ii) The strongest negative PTac-PTech correlation is between the \texttt{Threat of Loss} PTac and {\sf Incentive \& Motivator} PTech with $r = -0.83$, because the former uses loss as a threat and the latter uses gain as incentive (i.e., they are not compatible).

\begin{insight}
There are strong positive PTac-PTac and PTac-PTech correlations, but also not-so strong positive PTech-PTech correlations. Future defenses should strive to deal with the PTacs and PTechs that are often exploited together. 
\end{insight}

\noindent{\bf Addressing RQ4: What PTacs often exploit which PTechs, and what PTechs often exploit which PFs?}
Figure \ref{fig:mapping_PTac} shows the relationships between the PTacs and PTechs in the malicious emails. 
We determine which PTac exploits which PTech based on their definitions.
We say PTac $A$ exploits PTech $B$ if PTech $B$ is semantically related to PTac $A$. Thus, if a malicious email exploits PTech $B$, then the email also exploits PTac $A$, when is semantically related to PTech B. See concrete examples below.
We observe that all 7 PTacs exploit multiple PTechs. For example, the \texttt{Familiarity} PTac exploits the following 7 PTechs: (i) {\sf Persuasion}, by using the Principles of Persuasion (e.g., when the Commitment Principle is employed in an email together with a brand name that is familiar to a recipient); (ii) {\sf Impersonation}, by impersonating a familiar authority (e.g., one's boss);
(iii) {\sf Visual Deception}, by using visuals, such as logos, of popular brands to deceive a recipient;
(iv) {\sf Incentive \& Motivator} by using well-known or brand goods as an incentive (e.g., Walmart gift card); (v) {\sf Urgency}, by leveraging familiar situations, such as the Covid-19 infection, to encourage quick action; (vi) {\sf Attention Grabbing}, by using symbols that are familiar to a recipient to draw the recipient's attention;
and (vii) {\sf Contextualization}, by using common societal events that are familiar to a recipient (e.g., using the War in Ukraine to ask for donation).  
As another example, the \texttt{Threat to Identity} PTac exploits the following 2 PTechs: (i) {\sf Impersonation}, by assuming an entity that is tied to one's identity (e.g., ``Your memberships expires in 48 hours, please pay your dues to keep your membership"; and (ii) {\sf Personalization}, by addressing the recipient by name (for example) threatening a recipient about pictures that the attacker will send to the wife if the recipient does not give in to the demands of the attacker.

\ignore{ 

As another example, the \texttt{Threat of Loss} PTac exploits the following PTechs {\color{red}according to the dataset}: (i) {\sf Persuasion}, by using the principle of authority to threaten the recipient to act or lose something (e.g., ``Per our password renewal policy, you will lose your email storage if you do not update your password, Click on the link below to update now"); (ii) {\sf Impersonation} by pretending to be an entity with authority and asking the recipient to take an action or the recipient will lose something (e.g., ``Your utility bill is past due, pay immediately to avoid suspension of power") (iii) {\sf Urgency} by using time constraints to force the recipient to act quickly without thoughtfulness (e.g., ``Pay now or your utility will be suspended"), (iii) {\sf Personalization}, by calling the recipient by name before presenting a narrative, which the recipient must act upon or lose something (e.g., ``Hello Mark, your bill is due at midnight. Please, click here to pay so that you do not lose your account").
}

\begin{insight}
While all PTacs have exploited multiple PTechs, the \texttt{Fit \& Form} PTac has exploited all 9 PTechs.
\end{insight}

\vspace{-2.5em}
\begin{figure}[!htbp] 
\centering 
\includegraphics[width =\textwidth]{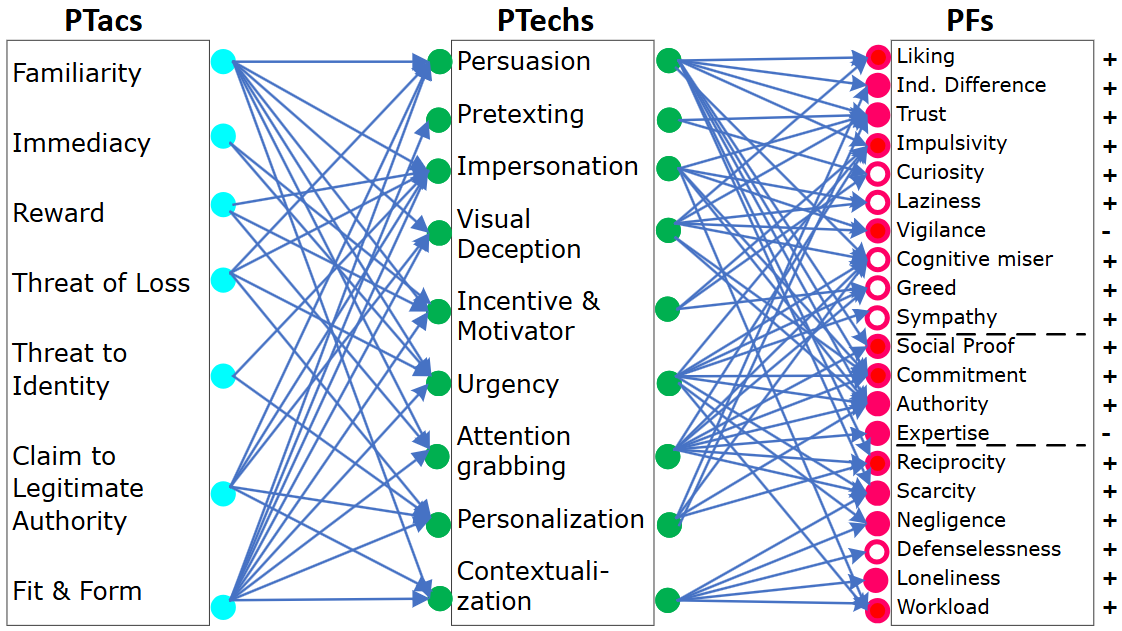}
\vspace{-1.5em}
\caption{\small Mappings of relationships between PTacs, PTechs, and PFs, where ``$A\to B$'' indicates ``$A$ exploits $B$,'' a dashed line inside a box indicates different categories, a filled circle indicates PFs with empirical quantitative studies and an empty circle otherwise, the ``+'' (``-'') sign indicates that a factors increases (decreases) human susceptibility to attacks \cite{longtchi2024internet}. ``Ind.'' is short for  Individual.}
\label{fig:mapping_PTac}
\end{figure}

Figure \ref{fig:mapping_PTac} also highlights the PTechs exploit PFs.
We observe that all 9 PTechs exploit multiple PFs (out of the 20 PFs specified in \cite{PF-paper}). For example, the {\sf Urgency} PTech exploits the following 9 PFs: (i) the {\sc Impulsivity} PF, by using time constraint elements in a malicious email (e.g. `` Click here to get your free copy"); (ii) the {\sc Cognitive miser} PF, by leveraging elements that give the recipient no time for any thoughtful decision (e.g., ``Free membership ends today, claim your free membership now"); (iii) the {\sc Greed} PF, by using elevated rewards and gains to bait greediness or the desire to get more (e.g., ``...6 million US dollars, and you will get 30\%"); (iv) the {\sc Sympathy} PF, by using a pathetic narrative that needs immediate action (e.g., ``Thieves stole all my belongings in an Airbnb, please send me \$1000 through Western Union. I will pay you back when I return");  (v) the {\sc Commitment} PF, by using what the recipients may be dedicated to, or doing with consistency (e.g., ``Please, donate now to support your party..."); (vi) the {\sc Authority} PF, by impersonating someone with some power over the recipient (e.g., emails purported to come from one's boss, saying ``Fill the attached form and send it to me urgently"); (vii) the {\sc Scarcity} PF, by saying a good or service is limited (e.g., ``Only few remaining, get yours now:); (viii) the {\sc Negligence} PF, by leveraging a situation that the recipient must be negligent; and (ix) the {\sc Workload} PF, by leveraging situations where the recipient is overwhelmed with work. 
Note that the last two PFs are opportunistic, because the attacker can only hope that the email will find the recipient in such situations (e.g., ``Review the attached file and send it to me asap"). 

Similarly, the {\sf Impersonation} PTech exploits 6 PFs, namely {\sc Trust}, {\sc Laziness}, {\sc Vigilance}, {\sc Commitment}, {\sc Authority}, and {\sc Workload}. For example, the {\sc trust} PF is exploited by assuming the personality of a trusted entity via the logo of the World Health Organization where the message reads, ``\textit{Download the attached Covid-19 treatment update}". 
The {\sf Contextualization} PTech exploits 5 PFs, namely {\sc Scarcity}, {\sc Negligence}, {\sc Defenselessness}, {\sc Loneliness}, and {\sc Workload}. For example, it exploits the {\sc scarcity} PF by using elements that indicate a fewer than normal in the the availability of goods and services, such as in the email message, ``\textit{Sign up here to be among the first people to get a free Covid-19 home test. Supply is limited}.'' 

\begin{insight}
The {\sf Attention Grabbing} PTech exploits most PFs, while the {\sc impulsivity} PF is most exploited by PTechs. 
\end{insight}

\section{Limitations}\label{sec:limitations}

This study has two limitations that need to be addressed in the future. First, the empirical study is based on one investigator because it demands a substantial amount of time and expertise in identifying PTacs and PTechs. This means that the results may be biased. However, our methodology is equally applicable when there are multiple  ``graders'' for identifying PTacs and PTechs from malicious emails, but would need to be extended with a calibration process, as shown in \cite{montanez2023quantifying}. 
Second, our study is based on 1,260 malicious emails over 21 years, or 60 emails per year. Thus, the results may not be as representative as desired. It would be ideal to analyze a much larger dataset (e.g., 60 emails per month).

\section{Conclusion}\label{sec:conclusion}

We have presented a methodology and applied it to a case study on characterizing the evolution of PTacs (Psychological Tactics) and PTechs (Psychological Techniques) exploited by malicious emails. We have drawn a number of insights, such as which PTacs and PTechs are exploited more often than others and which PTacs and/or PTechs are often exploited together. However, we have not observed the desired phenomenon that defenses have forced attackers to change their PTacs and/or PTechs because they become ineffective. 

The insights and the limitations of the present study shed light on future research directions toward designing effective defenses against malicious emails. Moreover, it is important to characterize the evolution of PTechs and PTacs that are exploited by the other kinds of attacks, such as malicious websites (e.g., \cite{MirIEEECNS2022,MirIEEEISI2020Landsapce,MirIEEEISI2020MalciousWebsites,XuCNS2014,XuCodaspy13-maliciousURL}) and online social engineering attacks. It is interesting to model and forecast the evolution of PTechs and PTacs, perhaps together with the evolution of PFs, in the same fashion as in 
\cite{XuSciSec2023-forecasting,XuTIFSSparsity2021,XuGrangerCausality2020,DBLP:journals/ejisec/FangXXZ19,XuJAS2018,XuTIFSDataBreach2018,XuJAS2016,XuTechnometrics2017,XuIEEETIFS2015,XuIEEETIFS2013} so as to enable proactive defense. It is also interesting to systematically define metrics to quantify the susceptibility of humans to cyber social engineering attacks in principled fashion \cite{Pendleton16,Cho16-milcom,XuSTRAM2018ACMCSUR,XuIEEETIFS2018-groundtruth,XuAgility2019,XuSciSec2021SARR}. 

\smallskip

\noindent{\bf Acknowledgement}. We thank the reviewers for their comments. This research was supported in part by NSF Grant \#2115134 and Colorado State Bill 18-086. 

\bibliographystyle{splncs04}
\bibliography{main}

\ignore{

}

\ignore{
\newpage
\begin{appendix}
    
\end{appendix}
}

\end{document}